\documentclass[fleqn,usenatbib]{mnras}
\usepackage{newtxtext,newtxmath}
\usepackage[T1]{fontenc}
\usepackage{ae,aecompl}
\usepackage{booktabs,makecell,multirow}

\usepackage{graphicx}	
\usepackage{amsmath}	

\usepackage{graphicx}
\usepackage{subfigure}
\newcommand\msun{{\,M_\odot}}

\newcommand\zsun{{\rm \,Z_\odot}}
\newcommand\lsun{{\rm \,L_\odot}}

\usepackage{aecompl}
\usepackage[T1]{fontenc}
\usepackage{threeparttable}
\usepackage{caption}
\usepackage{times}
\usepackage{ifpdf}

\newcommand{\cMpc}{~\mbox{comoving}~\mbox{Mpc}}

\newcommand{\cm}{~\mbox{cm}}

\title[r-process elements in ultra-faint dwarfs]{Highly r-process enhanced stars in ultra-faint dwarf galaxies}

\author[Jeon et al.]{Myoungwon Jeon$^{1}$\thanks{E-mail:
myjeon@khu.ac.kr}, 
Gurtina Besla$^{2}$, and Volker Bromm$^{3}$\\
$^{1}$School of Space Research, Kyung Hee University, 1732 Deogyeong-daero, Yongin-si, Gyeonggi-do 17104, Korea\\
$^{2}$Department of Astronomy, University of Arizona, 933 North Cherry Avenue, Tucson,
 AZ 85721, USA\\
$^{3}$Department of Astronomy, University of Texas, Austin, TX 78712, USA
}

\date{Accepted XXX. Received YYY; in original form ZZZ}

\pubyear{2019}

\begin{document}
\label{firstpage}
\pagerange{\pageref{firstpage}--\pageref{lastpage}}
\maketitle

\begin{abstract}
Highly r-process enhanced metal-poor stars (MP r-II, $\rm [Eu/Fe]>1$ and $\rm [Fe/H]\lesssim-1.5$) have been observed in ultra-faint dwarf (UFD) galaxy, specifically in Reticulum~II (Ret~II). The fact that only a few UFDs contain such stars implies that the r-process site may reflect very rare, but individually prolific events, such as neutron star mergers (NSMs). Considering the relatively short star formation history (SFH) of UFDs, it is puzzling how they could experience such a rare phenomenon. In this work, we show the results of cosmological hydrodynamic zoom-in simulations of isolated UFDs ($M_{\rm vir}\approx10^7-10^8\msun$ and $M_{\ast}\approx10^3-10^4\msun$ at $z=0$) to explain the formation of MP r-II stars in UFDs. We employ a simple toy model for NSM events, adopting parameters consistent with observations, such as the NSM rate (1 per $M_{\ast}\approx10^5\msun$) and europium (Eu) mass ($M_{\rm Eu}\approx10^{-5}\msun$). We identify only one simulated galaxy ($\rm M_{vir}\approx4.6\times10^7\msun$, $M_{\ast}\approx 3.4\times 10^3\msun$ at $z=0$) with abundances similar to Ret~II in a simulation volume that hosts $\sim30$ UFD analogs, indicating that such abundances are possible but rare. By exploring a range of key parameters, we demonstrate that the most important factor in determining the formation of MP r-II stars in UFDs is how quickly subsequent stars can be formed out of r-process enriched gas. We find that it takes between 10 to 100~Myr to form the first and second burst of MP r-II stars. Over this period, Eu-polluted gas maintains the required high abundance ratios of $\rm [Eu/Fe]>1$.

\end{abstract}

\begin{keywords}
cosmology: theory -- galaxies: formation -- galaxies: high-redshift -- HII regions --
hydrodynamics -- intergalactic medium -- supernovae: physics.
\end{keywords}



\section{Introduction}

Studying the formation and evolution of low-mass dwarf galaxies is essential to a comprehensive understanding of galaxy formation in $\Lambda$CDM cosmology \citep{Bullock2017, Strigari2018, Bromm2011}. Over the preceding decade, surveys of ever increasing size have assembled an array of observational constraints to understand such small systems, including the extreme end of the mass spectrum, ultra-faint dwarfs (UFDs), defined as galaxies with absolute magnitudes fainter than $M_{V}=-7.7$ $(L\lesssim10^5\lsun$; reviewed in \citealt{Simon2019}; see also \citealp{Tolstoy2009, Willman2010, McConnachie2012, Frebel2015}). Given their faintness, together with the paucity of metals in stellar components, it is inferred that they had experienced a relatively short star formation history (SFH) in the early Universe. Furthermore, it appears that select Milky Way (MW) UFDs had formed more than 70\% of stars prior to reionization, and then experienced quenching both by supernovae (SNe) and the global heating during reionization (e.g. \citealp{Brown2014, Weisz2014}).

A hint about the short SFHs of the MW UFDs is also imprinted in the abundance patterns of their member stars. For instance, the position of the knee in the $\rm [\alpha/Fe]$-$\rm [Fe/H]$ relation indicates how long star formation might last in UFDs, given that the $\alpha$-to-Fe ratio decreases as the amount of iron released over time by Type~Ia SNe increases (\citealp{Tinsley1979}). For UFDs, the location of the knee where $\rm [\alpha/Fe]$ begins to decline tends to be at a lower metallicity, $\rm [Fe/H]\lesssim-1$, indicating a shorter SFH \citep[][]{Frebel2012}. Moreover, the unique chemical signature in carbon-enhanced metal poor (CEMP) stars ($\rm [C/Fe]\gtrsim0.7$, $\rm [Fe/H]<-1$) implies that star formation in UFDs might predominantly take place in the early Universe (see, e.g. \citealp{Nomoto2013} for a review; see also \citealp{Yoon2019,Jeon2021}). The reason is that one of the promising pathways of producing the CEMP signature is connected to nucleosynthesis in the first generation of so-called Population~III (Pop~III) stars (e.g. \citealp{Bromm2013, Jeon2019}).

Considering their short SFHs, one of the challenges is to understand how UFDs with (abundant) neutron-capture elements, such as europium (Eu) and barium (Ba), can exist, given the rare occurrence of their production sites. Furthermore, stars with unusually high abundances of elements arising in the rapid neutron capture process ("r-process") are observed in select UFDs, specifically the Reticulum~II (Ret~II) galaxy (\citealp{Ji2016a, Ji2016b, Roederer2016}) with MP r-II stars ($\rm [Eu/Fe] > 1, [Fe/H] < -1.5$), and the Tucana~III (Tuc~III) system (\citealp{Hansen2017, Marshall2019}), which contains stars with a lower level of r-process enhancement (r-I, $\rm 0.3 < [Eu/Fe] < 1,  [Fe/H] < -1.5$) \citep{Beers2005}.

The origin of r-process elements in UFDs is still uncertain (e.g. \citealp{Argast2004, Beniamini2018}). A plausible pathway of producing r-process elements is via nucleosynthesis in neutrino-driven winds from core-collapse SN (CCSN) explosions (e.g. \citealp{Takahashi1994, Woosley1994, Arcones2013} for a review), a magneto-rotationally driven SN (e.g. \citealp{Nishimura2015}), or neutron star merger (NSM) events (e.g. \citealp{Freiburghaus1999, Wanajo2014}). Among them, the NSM site is the only one that is empirically confirmed, based on the recent multi-messenger observations of the GW170817 gravitational wave event (e.g. \citealp{Abbott2017, Chornock2017, Pian2017}), including the detection of a short-duration $\gamma$-ray burst (SGRB) electromagnetic counterpart  (e.g. \citealp{Goldstein2017, Savchenko2017}). Also, it is suggested that all r-process nuclides can be synthesized during NSM events (e.g. \citealp{Wanajo2014, Tsujimoto2014, Wu2016}), making them the most favored site of producing r-process elements.

Provided that only about 10\% of the observed UFDs appear to contain highly r-process enhanced stars (e.g. \citealp{Ji2016a}), it seems evident that the responsible event is rare but individually prolific, such as NSMs. Thus assuming that such a merger is the most promising site for their r-process elements, the question arises whether it is possible for UFDs to have experienced such rare events within their short SFHs ($<$1 Gyr). It is still unclear how often NSM events take place. Applying a maximum likelihood analysis to the observed dwarf spheroidal galaxies (dSphs) and UFDs, \citet{Beniamini2016} have suggested that the expected NSM rate is $5\times 10^{-4}-2\times10^{-3}$ per CCSN, which corresponds to a single NSM event per $500-2000$ CCSNe. Adopting a typical stellar initial mass function (IMF), about $500-2000$ CCSNe can be hosted in a galaxy with a stellar mass of $M_{\ast}\approx10^5\msun$ (\citealp{Tsujimoto2015}). Given that the stellar masses of Ret~II and Tuc~III are $M_{\ast}=2.6\pm0.2\times10^3\msun$ (\citealp{Bechtol2015}) and $M_{\ast}=0.8\pm0.1\times10^3\msun$ (\citealp{Drlica2015}), respectively, such UFDs are less likely to host even a single NSM event during their SFHs.

Another challenge is that some stars in select UFDs show significantly high r-process element ratios ($\rm [Eu/Fe]\geq1.0, [Eu/H]\geq-1.0$). In general, depending on the $\rm [Eu/Fe]$ ratio, r-process enhanced metal-poor (MP) stars are conventionally divided into two subclasses, MP~r-I ($\rm 0.3 < [Eu/Fe] < 1,  [Fe/H] < -1.5$) and MP~r-II ($\rm [Eu/Fe] > 1, [Fe/H] < -1.5$) \citep{Beers2005}. In particular, in Ret~II, seven out of nine stars show extremely high levels of r-process element ratios, $\rm [Eu/Fe]>1.7$ (e.g. \citealp{Ji2016b}). This implies that once r-process elements are released, subsequent star formation should occur promptly out of the r-process enriched gas clouds. Otherwise, the ejected r-process elements might be diluted as they mix with the surrounding medium, giving rise to a low $\rm [Eu/H]$ ratio. By understanding how such high [Eu/Fe] ratios can be achieved, one can infer the physical mechanisms of the r-process. For instance, assuming that the NSM pathway is the most compelling site for synthesizing the observed r-process elements, we can infer how frequently such mergers might occur or how efficiently the r-process elements can mix into the interstellar medium (ISM). In this work, we specifically consider the r-process from NSM events to explain the observed r-process elements in select UFDs.

Here we present a new suite of hydrodynamic zoom-in simulations, where a few tens of UFD analogs with $M_{\rm vir}\approx10^7-10^8\msun$ (at $z=0$) form. We will investigate whether the simulations can reproduce the significant Eu enhancement found in the MP~r-II stars in Ret~II. To do so, we employ two approaches. First, we study if r-process enhanced stars can self-consistently be formed, when using the NSM rate and metal yields constrained by observations. Secondly, we explore the conditions under which MP~r-II stars can be created by changing the relevant parameters in an idealized setting. Note that, in this work, we consider only europium, treating it as a representative r-process element. This paper is organized as follows. In Section~2, we explain the numerical methodology, followed by a presentation of the detailed simulation results in Section~3, with a focus on the formation of r-II stars in the context of galaxy assembly. We conclude in Section~4 with a summary. For consistency, all distances are given in physical (proper) units unless noted otherwise.

\begin{table*}
\caption{Summary of the simulations. 
Column (1): Simulation name. In two relatively large volumes, {\sc LV1} and {\sc LV2}, a few tens of UFD analogs are simultaneously formed. Meanwhile, in {\sc Single galaxy} we have only focused on a single galaxy by changing the NSM parameters. Column (2): Number of UFD analogs in the simulated volume. Column (3): Run name. We have conducted comparison runs by increasing the NSM rate by a factor of 2 ({\sc Fiducial-2x}) and 3 ({\sc Fiducial-3x}), and decreasing Eu yield in half ({\sc Fiducial-2x-half}). Column (4): NSM rate per $\msun$. 
Column (5): Eu yield in $\msun$ from a single NSM event. Column (6): Presence of high [Eu/H] stars, formed in the run.}
\setlength{\tabcolsep}{5.2mm}{
\begin{tabular}{cccccc}
\toprule
\makecell[c]{Name} &\makecell[c]{ Number of UFDs}& \makecell[c]{Run name}  &\makecell[c]{NSM rate} &\makecell[c]{Eu yield [$\msun]$} &\makecell[c]{r~II stars}
\\
\midrule
\multirowcell{1}{{\sc LV1}}&8&-&1 per $1\times10^5\msun$& $4.5\times10^{-5}$& No\\

\midrule
\multirowcell{1}{{\sc LV2}}&32&-&1 per $1\times10^5\msun$& $4.5\times10^{-5}$& Yes\\

\midrule
\multirowcell{5}{Single galaxy}&1& {\sc fiducial} &1 per $3.5\times10^3\msun$& $4.5\times10^{-5}$& No\\
&1& {\sc fiducial-2x} &1 per $7.0\times10^3\msun$& $4.5\times10^{-5}$& Yes\\
&1& {\sc fiducial-2x-half} &1 per $7.0\times10^3\msun$& $2.0\times10^{-5}$& Yes\\
&1& {\sc fiducial-3x} &1 per $1.0\times10^4\msun$& $4.5\times10^{-5}$& No\\

\bottomrule
\end{tabular}}
\end{table*}

\section{Numerical methodology}
\label{Sec:Metho}

\subsection{Simulation Setup}
The simulations have been carried out using a modified version of the N-body Smoothed Particle Hydrodynamics (SPH) code {\sc GADGET} (\citealp{Springel2001}; \citealp{Springel2005}) implemented with well-tested modules for baryon physics, mainly taken and modified from the OWLS simulations (\citealp{Schaye2010}). We generate cosmological initial conditions using {\sc MUSIC} (\citealp{Hahn2011}). As mentioned above, we have performed a suite of simulations in two settings. 
\\
\begin{itemize}
\item {\sc LV1, LV2:} we employ a simulation set-up such that a single NSM event might occur per a few tens of UFDs, as inferred from observations (\citealp{Ji2016a}). We have chosen two relatively large volumes with a box size of $L=3.125 h^{-1} \cMpc$ ({\sc LV1}) and $L=4.5 h^{-1} \cMpc$ ({\sc LV2}), respectively, to contain multiple UFD analogs with haloes in the mass range of $M_{\rm vir}\sim10^7-10^8\msun$ at $z=0$. The goal of the setting is to see whether MP r~II stars can self-consistently be formed under conditions similar to observations.
\\
\item {\sc Single galaxy:} we focus on a single fiducial galaxy ($M_{\rm vir}\sim10^8\msun$ at $z=0$) to explore parameters associated with the NSM model, such as the merger rate and europium yield. Here, we aim to study the necessary conditions for the formation of MP r~II stars by changing the involved parameters in an idealized setting.
\end{itemize}
Four consecutive refinements are conducted in all volumes, resulting in an effective resolution of $2048^3$, giving rise to mass resolutions of $m_{\rm DM} \approx 500\msun$ and $m_{\rm SPH} \approx 65\msun$, respectively, for dark matter and gas in the most refined region. The basic physics in these simulations is described in \citet{Jeon2021}, and thus we refer the reader to this paper for details, but briefly summarize key aspects here.

\subsection{Baryon physics}

We include all relevant primordial cooling processes and metal cooling to trace the evolution of atomic and molecular species ($\rm H, H^{+}, H^{-}, H_{2}, H^{+}_2 , He, He^{+}, He^{++}, e^{-}, D, D^{+}$, and HD) by solving the coupled, non-equilibrium rate equations. Once the gas density exceeds a density threshold of $n_{\rm th}=100\cm^{-3}$, a gas particle is converted into a collisionless star particle. Stars arise from gas clouds according to Schmidt's Law (\citealp{Schmidt1959}), with a star formation efficiency per free fall time for both Pop~III and Pop~II stars of $\epsilon_{\rm ff}=0.01$. We implement the two stellar populations differently. Since first generation stars are likely to be massive and thus can be resolved in this work, they are randomly sampled as individual star from an assumed top-heavy IMF (\citealp{Chabrier2003, Wise2012}). Meanwhile, owing to their small masses, Pop~II stars are described as a stellar cluster, composed of a single stellar population, with a mass of $M_{\rm \ast, Pop~II}=500\msun$, following a Salpeter IMF. The transition from Pop~III to Pop~II star formation mode is achieved when the gas metallicity is larger than the critical metallicity of $Z_{\rm crit}=10^{-5.5}\zsun$ (e.g. \citealp{Omukai2000}; \citealp{Bromm2001a}).

When a star dies, it undergoes a SN explosion with properties that depend on the mass of the star. For Pop~III stars, we consider conventional CCSNe and powerful pair-instability SNe (PISNe), and Type~II and Type~Ia SNe for Pop~II stars. The resulting SN energy is released as thermal energy into the surrounding medium. Metal yields and total ejected masses for Pop~III stars are drawn from \citet{Heger2010} for CCSNe, arising from Pop~III progenitors in the mass range of $m_{\ast, \rm Pop~III}=10-100\msun$, and from \citet{Heger2002} for PISNe, which are expected for massive stars of $m_{\ast \rm Pop~III}=140-260\msun$. For Pop~II stars, metals from asymptotic giant branch (AGB) enrichment (\citealp{Marigo2001}), Type~II (\citealp{Portinari1998}, and Type~Ia SNe (\citealp{Thielemann2003}) are included. 

Initially, the metals are released onto the neighboring gas particles, $N_{\rm ngb}=48$, and then they mix with the surrounding ISM via a diffusion process (\citealp{Greif2009}). Details can be found in \citet{Jeon2017}, but for the convenience of the reader, we here summarize the key elements of our numerical approach to treat the transport of metals. Specifically, the transport process is modelled by solving the diffusion equation \begin{equation}
\frac{dc}{dt} = \frac{1}{\rho} \nabla \cdot (D \nabla c)\mbox{\ ,}
\end{equation}
where $c$ is the concentration of a quantity of interest. In this work, it corresponds to the total gas metallicity or an individual metal species. How efficiently metals can be diluted is determined by the diffusion coefficient (\citealp{Klessen2003}), defined as $D=2 \rho v l$, where $l$ is the characteristic scale for turbulent driving, $\rho$ the gas density, and $v$ the velocity dispersion in the surrounding medium.

In addition to SN feedback, we impose heating from a global UV background (UVB) (\citealp{Haardt2011}), starting at $z\approx7$ and gradually increasing to the maximum strength at $z\approx6$, corresponding to the era where reionization is complete (e.g. \citealp{Gunn1965}; \citealp{Fan2007}). Also, we account for the photodissociation of molecular hydrogen, $\rm H_2$, by soft UV radiation in the Lyman-Werner (LW) band (11.2 eV$-$13.6 eV) (e.g. \citealp{Abel1997}). We, however, should note that the photoionization heating from individual Pop~III stars or Pop~II clusters is not considered, which could suppress star formation in the simulated galaxies. We will discuss the impact of the photoionization heating on the SFHs of UFD analogs in an upcoming paper.

\begin{figure*}
  \includegraphics[width=130mm]{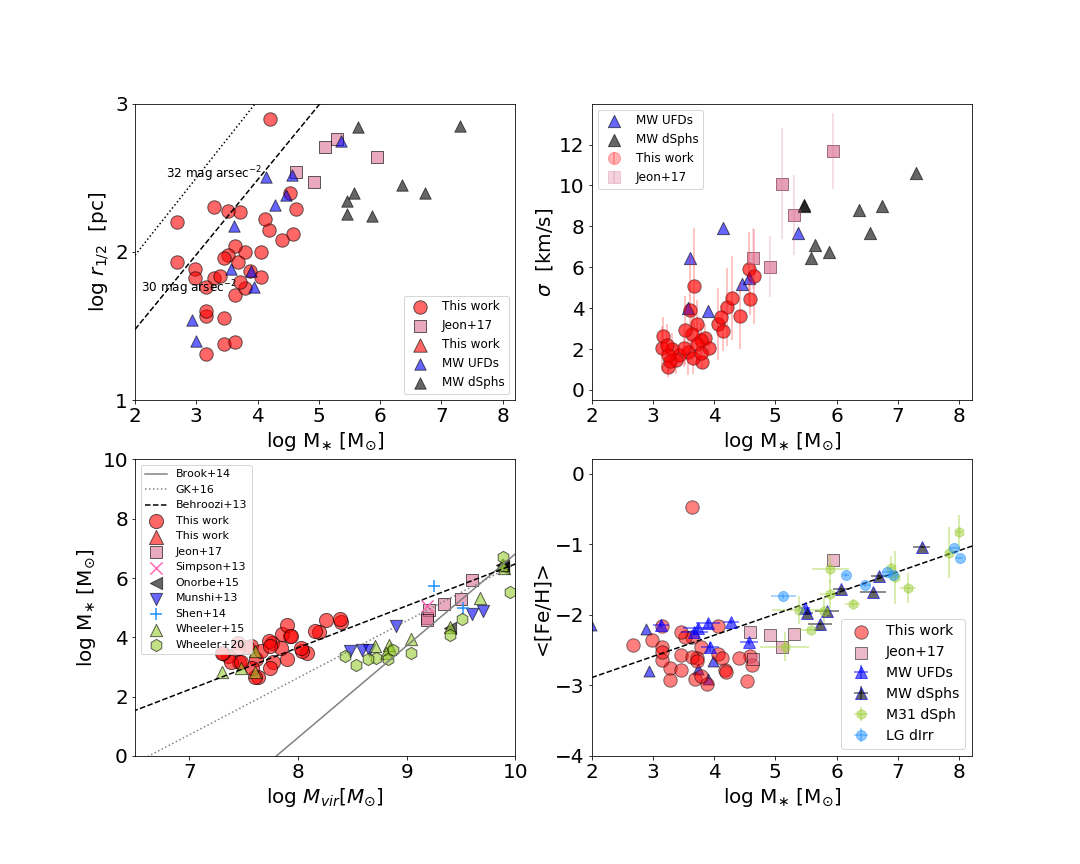}
   \caption{Physical quantities derived from the simulated UFDs at $z=0$. Each panel (clockwise from upper left) presents half-stellar mass radius, average velocity dispersion of stars along the line of sight, average stellar metallicity as a function of stellar mass, and the halo-stellar mass relation. The estimated values from this work are shown as red filled circles, by comparing with those of the MW UFDs (blue triangles) and dSphs (black triangles) (e.g. \citealp{Kirby2013, Simon2019}). We note that the resulting properties, especially the half-stellar mass radius and average metallicity, seem to show good agreement with what is observed. Regarding the halo-stellar mass relation (lower left panel), our result is consistent with the slope (dashed line) provided by \citet{Behroozi2013}, whose fit is based on the abundance matching technique.}
\end{figure*}

\subsection{Neutron-star mergers}

\subsubsection{Merger rate}
In the runs {\sc LV1} and {\sc LV2}, we model NSM events with a fiducial rate of one merger per $10^5\msun$ in stars, consistent with the average expected rate for such mergers (e.g. \citealp{Tsujimoto2015}). Assuming an empirical stellar IMF, one NSM is expected per 1000-2000 CCSNe, corresponding to a stellar mass of $\sim10^5\msun$. This estimate is consistent with the rate inferred from the observed UFDs \citep{Ji2016a}, where it is suggested that a single NSM event might occur in the sample of ten UFDs. Such tens of UFDs would host $\sim$2000 SNe in total. We note, however, that the canonical NSM rate can be boosted by up to a factor of 10, due to uncertainties in the underlying population synthesis modeling (\citealp{Dominik2012}).

For the fiducial single galaxy run ({\sc fiducial}), we adopt a rate of one NSM per $3.5\times10^3\msun$, an order of magnitude larger than the canonical value, to ensure that at least one NSM event happens in the SFH of our simulated galaxies, in line with the stellar mass of Ret~II of only $M_{\ast}\approx5\times10^3\msun$. To assess the impact of model uncertainty, we carry out comparison simulations:
\begin{itemize}
    \item {\sc fiducial:} we adopt a rate of one NSM per $3.5\times10^3\msun$ and the europium yield of $M_{\rm Eu}\approx4.5\times10^{-5}\msun$.
    \item {\sc fiducial-2x:} we decrease the NSM rate by a factor of 2, such that one NSM event takes place per $7\times10^3\msun$.
    \item {\sc fiducial-3x:} we decrease the NSM rate by a factor of 3, such that one NSM event takes place per $1\times10^4\msun$.
    \item {\sc fiducial-2x-half:} we decrease the NSM rate by a factor of 2 and reduce the Eu yield by half, $M_{\rm Eu}\approx2\times10^{-5}\msun$.
\end{itemize}
The parameter ranges for those simulations are summarized in Table~1. We assume that the final merger occurs about 5~Myr after the binary neutron star system forms. Such delay time of $\sim5$~Myr adopted here corresponds to a fast-merging channel (e.g. \citealp{Belczynski2001, DeDonder2004, Dominik2012}). In addition, we ignore natal kicks and the explosion energy of the NSM event, which means that the estimated r-process element abundances in this work should be considered as upper limits.


\subsubsection{Nucleosynthetic yield}
We choose a fiducial value for the europium yield of $M_{\rm Eu}\approx4.5\times10^{-5}\msun$ for each NS merger event, adopted by \citet{Naiman2018}, as inferred by the chemical evolution of europium in MW-sized galaxies in the IllustrisTNG simulation. This value, however, is three times higher than the upper limit based on the NS-NS merger detected by LIGO/Virgo for the GW170817 event (\citealp{Cowperthwaite2017, Cote2018}). We use this high europium yield from \citet{Naiman2018} in order to assess whether the high europium abundance observed in Ret~II can be reproduced through simulations. In addition, we have conducted another simulation ({\sc Fiducial-2x-half}), where the Eu yield is reduced by half, more consistent with the value inferred from GW170817. We treat the ejection and transport of europium in the same way as other metal species from SN explosions. Once europium nuclei are ejected, they are initially distributed among the neighboring gas particles, followed by the subsequent diffusive transport according to our numerical scheme (see above for details).

\section{Simulation results}
In this section, we present the results for the simulated galaxies. First, we show the global properties to see if the simulated UFD analogs are in reasonable agreement with observational data. Second, we discuss the probability of forming a Ret~II-like system within the {\sc LV1} and {\sc LV2} runs. We then assess under which conditions extremely r-process enhanced stars can be formed using the "Single Galaxy" simulation suite and the variations in NSM rate and Eu yield described in section 2.3. Finally, we compare our results with other theoretical work regarding r-process elements in UFDs.

\begin{figure*}
  \includegraphics[width=160mm]{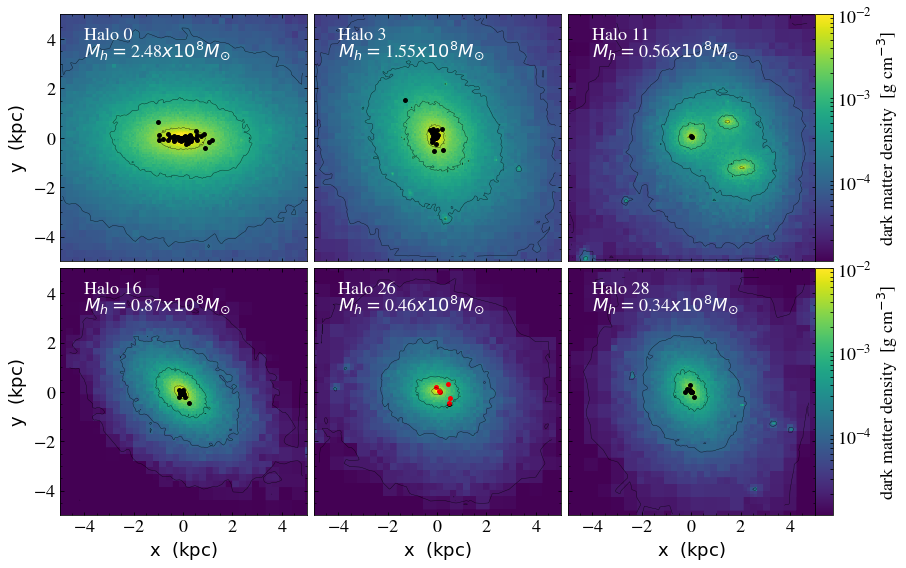}
   \caption{Projected dark matter profiles for select UFDs in {\sc LV2} at $z=0$. Pop~II clusters are shown as filled circles, with color the denoting Eu abundances, such that red circles indicate MP r-II stars ($\rm [Eu/Fe]>1$). All smaller values are presented as black circles. Out of 32 UFD analogs, we find only one halo with MP r-II stars ($M_{\rm vir}=4.6\times10^7\msun$, bottom middle panel). Despite adopting the same parameters for the NSM model, none of the galaxies in {\sc LV1}, where 8 UFDs arise, contain MP r-II stars. As such, the frequency of Ret~II-like systems predicted in this work (1 in a few tens of UFDs) is largely similar to what observations suggest.
   }
   
\end{figure*}

\begin{figure*}
  \includegraphics[width=120mm]{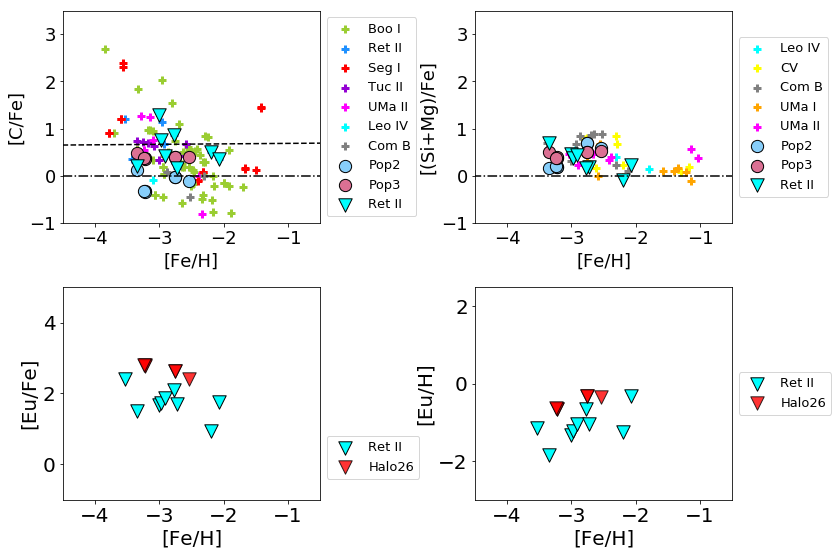}
   \caption{Stellar abundances of Ret~II-like systems (Halo~26). {\it Clockwise from upper left:} carbon-to-iron, $\alpha$-element-to-iron, absolute europium, and europium-to-iron ratio. The carbon and $\alpha$-element ratios are separately calculated with metals from Pop~III (pink circles) and Pop~II (blue circles) SNe, respectively. Note that no CEMP stars ($\rm [C/Fe]>0.7$) are found in the halo. We also add the chemical abundances of the MW UFDs, denoted as crosses with different colors. We identify 6 MP r-II stars (red inverse triangles) out of 7 Pop~II clusters. The resulting europium ratios show reasonably good agreement with the estimates in Ret~II.}
\end{figure*}

\begin{figure}
\centering
  \includegraphics[width=50mm]{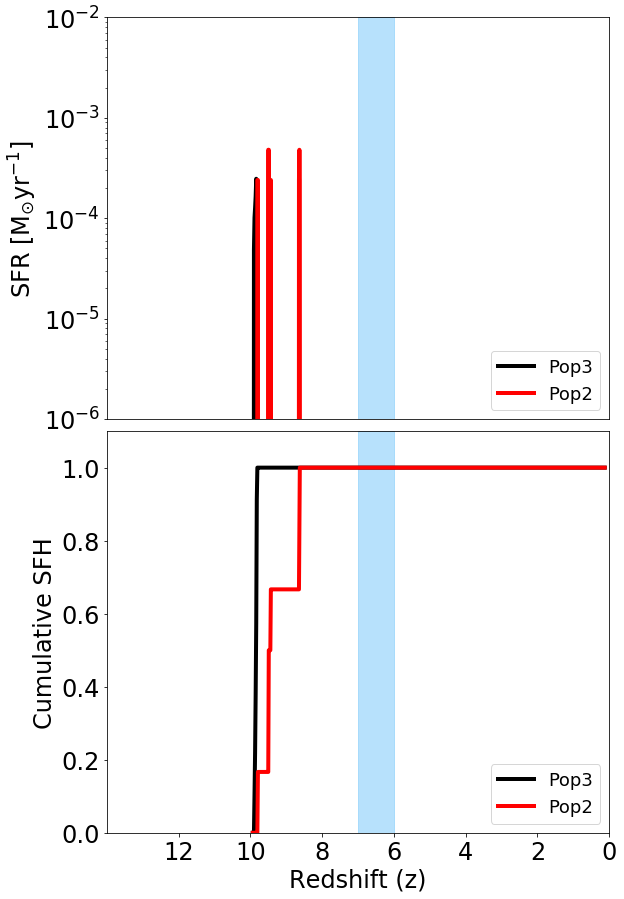}
   \caption{Star formation rate (top) and cumulative SFH (bottom) for Pop~III (black) and Pop~II (red) stars, formed in a Ret~II-like system ({\sc Halo 26}). The vertical shaded region indicates the epoch of reionization which starts at $z=7$, and its amplitude linearly increases up to $z=6$. Note that here, self-quenching is achieved by SN~feedback prior to reionization. In this halo, star formation takes place over $\sim$100 Myr.}
\end{figure}

\subsection{Global properties}

In Figure~1, we show the physical properties of the simulated galaxies at $z=0$ in {\sc LV2}. The panels (clockwise from upper left) present the half-stellar mass radius ($r_{\rm 1/2}$), average line of sight velocity dispersion ($\sigma$), and average stellar metallicity (<[Fe/H]>) as a function of stellar mass ($M_{\ast}$), as well as the stellar-halo mass relation. To compare the resulting quantities with observations, we display measurements for the MW UFDs (blue triangles) and dSphs (black triangles) (e.g. \citealp{McConnachie2012, Kirby2013, Simon2019}). Also, we include results of simulated galaxies (denoted as pink squares) from \citet{Jeon2017}, exploring systems an order of magnitude more massive than the UFD analogs in this work. We point out that the size ($r_{\rm 1/2}$) and the average metallicity (<[Fe/H>) of the simulated galaxies in the present work (red circles) agree well with those of the observed MW UFDs (e.g. \citealp{McConnachie2012}). 
 
We compute the stellar velocity dispersion of the simulated galaxies, $\sigma$, by averaging over the values along the 1000 random lines of sight. Note that we only consider the stars within the half-stellar mass radius to derive $\sigma$. The error bars arise from the variation of the velocity dispersion of 1000 random lines of sight. We find that most of the simulated galaxies below $M_{\ast}=10^4\msun$ tend to exhibit velocity dispersions of less than 4 km $\rm s^{-1}$, while the inferred values from observations range from 4 km $\rm s^{-1}$ to 8 km $\rm s^{-1}$ for the MW UFDs. Although the error bars encompass the observational data, we plan to conduct a follow-up study to understand such systematic discrepancy in future studies.

The halo-stellar mass relation is shown in comparison with other theoretical results, derived from high resolution zoom-in hydrodynamic simulations (\citealp{Munshi2013, Simpson2013, Shen2014, Onorbe2015, Wheeler2015, Jeon2017, Wheeler2019}), and large-box simulations in which best-fits are provided (\citealp{Brook2014} (solid line), \citealp{GK2016} (dotted line)). Our $\rm M_{\rm vir}$-$\rm M_{\ast}$ relation, in particular, appears to be in line with the fit suggested by \citet{Behroozi2013} (dashed line), whose result is based on the abundance matching approach. We should, however, caution that the effect of photoionization heating from local sources such as individual Pop~III stars and Pop~II clusters is not considered in this work. The impact of radiative feedback by local sources is still under debate. A recent study by \citet{Hopkins2020} has shown that the stellar mass of small galaxies ($M_{\rm vir}=2.4\times10^9 \msun$) is predominantly determined by the external UVB, but that the effect of local sources is insignificant. \citet{Agertz2020}, on the other hand, have demonstrated that the photoionization heating by stars could lower the stellar mass by a factor of $5-10$ for dwarf galaxies with a mass of $M_{\rm vir}=10^9 \msun$. This discrepancy is likely due to the difference in the detailed prescription for sub-grid models and resolution.

\begin{figure}
    \centering
    \subfigure{{\includegraphics[width=4.4cm]{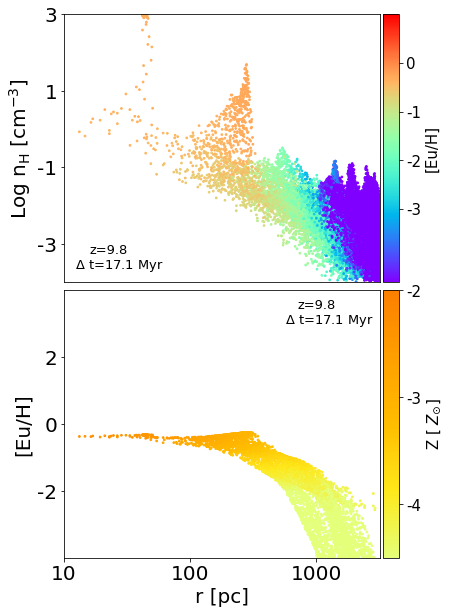} }}%
    \subfigure{{\includegraphics[width=4.4cm]{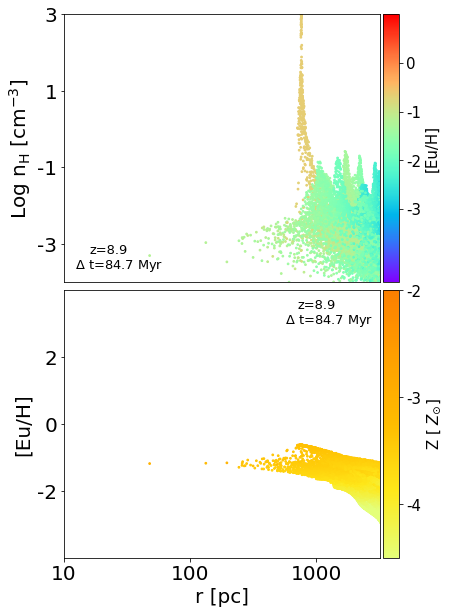} }}%
    \caption{Hydrogen number density (top panels) and Eu abundance (bottom panels) of the gas, as a function of distance from the NSM site at 17 (left) and 85 (right) Myr after the NSM event, corresponding to the first and second burst of star formation, respectively.}%
    \label{fig:example}%
\end{figure}

As shown in the bottom right panel of Figure~1, the resulting average metallicities (<[Fe/H]>) of the simulated galaxies tend to show a good agreement with those observed in UFDs, except one outlier having metallicity as high as $\rm [Fe/H]\approx-0.4$. We attribute such high metallicity system to Pop~III PISNe, which typically eject significantly more metals than CCSNe (e.g. \citealp{Karlsson2013, Heger2010}). We find that the high metallicity stars are formed at $z\approx9.3$, and about 80-90\% of their metals come from Pop~III stars. The metals, in particular, originate from a Pop~III PISN that explodes at $z\approx9.6$ within the progenitor halo.

\subsection{UFD analogs: r-process enhanced stars}
\subsubsection{Large volume runs: {\sc LV1} and {\sc LV2}}
 
As a first step, we consider whether any of the simulated UFDs in {\sc LV1}, in which 8 UFD analogs ($M_{\rm vir}\approx10^7-10^8\msun$) emerge, could host MP~r-II stars by adopting the empirical NSM rate (\citealp{Ji2016a}), where one NSM event is expected per $M_{\ast}\approx10^5\msun$. We further use the fiducial value for the europium yield, $M_{\rm Eu}\approx4.5\times10^{-5}\msun$. However, none of the simulated galaxies has stars with $\rm [Eu/Fe]>1$ and $[\rm Fe/H]<-1.5$, despite the occurrence of the NSM event in {\sc LV1}. We find that the simulated galaxies in {\sc LV1} continue to form stars after the NSM event, but the NSM ejecta is too diluted due to the long time delay between the NSM event and the subsequent star formation, resulting in the lack of MP~r-II stars.


Using the same empirical parameters used in {\sc LV1}, on the other hand, out of a total of 32 UFD analogs in {\sc LV2}, we identify a single galaxy ({\sc Halo~26}), containing MP r-II stars. The corresponding halo has virial and stellar masses of $\rm M_{vir}\approx4.6\times10^7\msun$ and $M_{\ast}\approx 3.4\times 10^3\msun$, respectively, at $z=0$. For illustrative purposes, in Figure~2, we show the projected dark matter profile of the representative UFD analogs at $z=0$ in {\sc LV2}. Pop~II stellar components are denoted as filled circles, with color indicating the $\rm [Eu/Fe]$ ratio such that MP r-II stars are shown in red. For further analysis, we only focus on {\sc Halo 26}, displayed in the bottom-middle panel of Figure~2, as it is the only simulated galaxy with abundances similar to Ret~II.

In Figure~3, we present stellar abundances (C, Si, Mg, and Eu), calculated from stars hosted in {\sc Halo~26}, by comparing with the observed values\footnote{The original references for the abundance ratios of the observed UFDs are as follows: \citealp{Norris2010, Lai2011} (Bootes~I); \citealp{Ji2016b, Roederer2016} (Reticulum~II); \citealp{Norris2010, Frebel2014} (Segue~I); \citealp{Ji2016c, Chiti2018a, Chiti2018b} (Tucana~II); \citealp{Frebel2010} (Ursa Minor~II); \citealp{Simon2010} (Leo~IV); \citealp{Frebel2010} (Coma Berenices)}. The first point to note here is that there are no CEMP stars ($\rm [C/Fe]\geq0.7$, $\rm [Fe/H]\leq-1$) in this galaxy. We compute the $\rm [C/Fe]$ ratio based on the metals separately released by Pop~III (pink circles in the top left panel of Fig.~3) and Pop~II (blue color), respectively (e.g. \citealp{Jeon2021}). There are three CEMP stars in Ret~II (\citealp{Ji2016b}), accounting for 40\% of stars at $\rm [Fe/H]\lesssim-3$. In general, Pop~III SNe are one of the favoured pathways of forming CEMPs (e.g. \citealp{Iwamoto2005, Heger2010, Ishigaki2014}). Although we include Pop~III stars, giving rise to a number of CEMP stars in the overall sample of UFD analogs in {\sc LV2}, no CEMP stars are found in {\sc Halo~26}. This deficiency can simply be understood as a stochastic effect (e.g. \citealp{Jeon2021}). Meanwhile, as shown in the upper-right panel of Figure~3, $\alpha$-element abundances of stars in {\sc Halo~26} are in good agreement with those of Ret~II.

The estimated europium abundances, $\rm [Eu/Fe]$ and $\rm [Eu/H]$, of the Pop~II clusters are presented in the bottom panels of Figure~3. Out of 7 Pop~II clusters, 6 (red inverse triangles) show extremely high europium abundances ($\rm [Eu/Fe]\gtrsim2.5, [Eu/H]>-0.6$), comparable to MP r-II stars in Ret~II. Despite the same assumptions for the NSM model in both {\sc LV1} and {\sc LV2}, Ret~II-like systems are produced only in {\sc LV2}. To better understand the formation of MP~r-II stars in the context of overall galaxy assembly, in Figure~4, we illustrate star formation rates (top panel) and cumulative SFH (bottom panel) for Pop~III (black solid lines) and Pop~II (red solid lines) stars in {\sc Halo26}, as a function of cosmic time. We find that in {\sc Halo26} Pop~III star formation begins at $z\approx10.3$, and the transition from Pop~III to Pop~II star formation mode occurs after $\sim$30 Myr. The heavier the halo, the earlier the star formation occurs, and the onset of star formation in the all simulated galaxies ranges from $z\approx13$ to $z\approx8$. {\sc Halo26} experiences self-quenching even before reionization, implying that the gas evaporated by SN~feedback could not recollapse onto the halo prior to reionization \citep{Jeon2014}. Note that the SFH of {\sc Halo 26} is not unusual, as about $25\%$ of the simulated galaxies tend to exhibit the self-quenching feature.

At $z\approx9.7$, a binary neutron star system forms, and after a time delay of 5~Myr we assume that the final merger takes place, ejecting r-process elements into the surrounding medium, out of which r-process enhanced Pop~II stars are formed. We will further discuss such a fast-merging channel below (see Section~3.3). It takes $\Delta t=17$~Myr to form MP r-II stars with $\rm [Eu/Fe]\approx2.6$ ($\rm [Eu/H]\approx-0.2$) since the release of r-process elements. Subsequently, a second burst of star formation occurs at $\Delta t=85$~Myr, creating MP r-II stars with $\rm [Eu/Fe]\approx2.7$ ($\rm [Eu/H]\approx-0.6$). To examine the gas conditions, when the MP r-II stars are formed, we show the hydrogen number density (top) and europium abundance (bottom) in Figure~5, as a function of distance from the NSM site at $\Delta t=17$ Myr (left) and $\Delta t=85$ Myr (right), corresponding to the first and second burst of star formation. 
\begin{figure}
  \includegraphics[width=90mm]{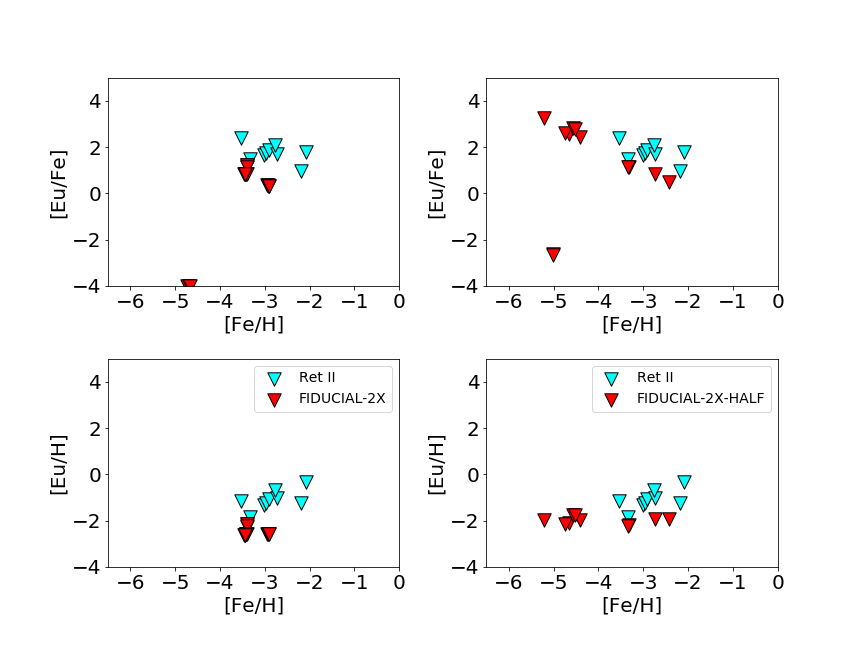}
  \caption{Europium abundances in the comparison runs, {\sc fiducial-2x} (left) and {\sc fiducial-2x-half} (right). We only show the results of simulations that contain highly r-process enhanced stars. To avoid confusion, we note that the derived abundances here are from {\sc Single galaxy} runs, while Figure~3 shows the stellar abundances of {\sc Halo26} in {\sc LV2}. {\it Top row:} [Eu/Fe] vs. [Fe/H] ratios. We show values derived from the simulations (red triangles) and compare them with those observed for r-process enhanced stars in Ret~II (cyan triangles). {\it Bottom row:} [Eu/H] vs. [Fe/H], with a similar convention for the symbols as above.}
\end{figure}
The NSM occurs at $z=9.7$ towards the center of {\sc Halo26}, when the virial mass and radius of the progenitor are  $M_{\rm vir}=4.3\times10^6\msun$ and $R_{\rm vir}=480$~pc, respectively. During the first period of suppressing star formation, lasting $\Delta t=17$ Myr, a gas mass of $M_{\rm gas}\approx4.3\times10^5\msun$ within the halo is polluted by europium up to {\bf $\rm [Eu/H]=-0.2$}. Out of the Eu-enriched density peak, located at $\sim44$~pc from the NSM site, as illustrated in the left-top panel of Figure~5, MP r-II stars are formed. The gas density inside the halo decreases mainly due to SN~feedback, and the europium is diluted beyond the virial radius. However, we find that the density peak at $\sim$550~pc still maintains a high europium abundances of {\bf $\rm [Eu/H]\approx-0.6$}, resulting in the formation of another MP r-II stars at $\Delta t=85$~Myr.

To assess how quickly the r-process elements are diluted, we calculate the diffusion coefficient, $D$, assuming a Gaussian distribution function for the transport of metals, with a standard deviation of $\sigma_{\rm D}$ (e.g. \citealp{Tarumi2020}). By fitting the relation, $2Dt = \sigma_{\rm D}^2$, the derived diffusion coefficient within {\sc Halo 26} is $D=2.3\times10^{-3}\,\rm kpc^{2}\, Myr^{-1}$ at $\Delta t=17$ Myr. This value agrees within a factor of $\sim 2$ with the estimates in \citet{Tarumi2020}, where $D=1\times10^{-3}\rm \,kpc^{2}\, Myr^{-1}$ is suggested inside a $M_{\rm vir}\approx10^8\msun$ halo at $z=8$. The impact of turbulent mixing on the r-process elements is also stressed in the work by \citet{Dvorkin2020}, where they reveal that metal mixing is essential to account for the large scatter in the [\rm Eu/Fe] ratio, especially for low $\rm [Fe/H]<-3$, observed in the MW stars. Recently, \citet{Beniamini2020} propose a new constraint on the diffusion coefficient, $D>10^{-4}\rm \,kpc^{2}\, Myr^{-1}$, assuming an r-process rate of $4\times10^{-5}\, \rm yr^{-1}$, based on MW stars using a Monte Carlo simulation and analytic approach.

\begin{figure*}
   \centering
    \subfigure{{\includegraphics[width=5.5cm]{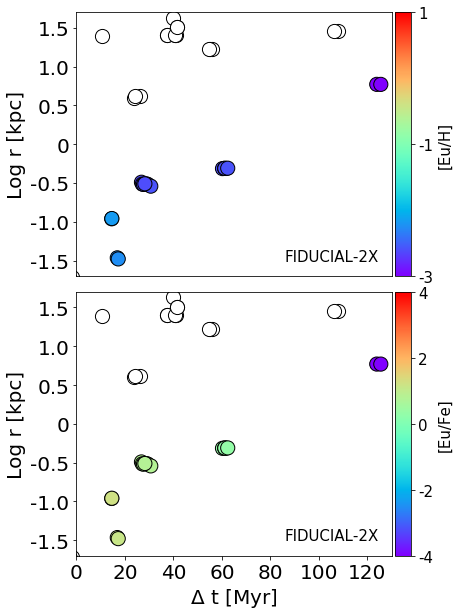} }}%
    \subfigure{{\includegraphics[width=5.5cm]{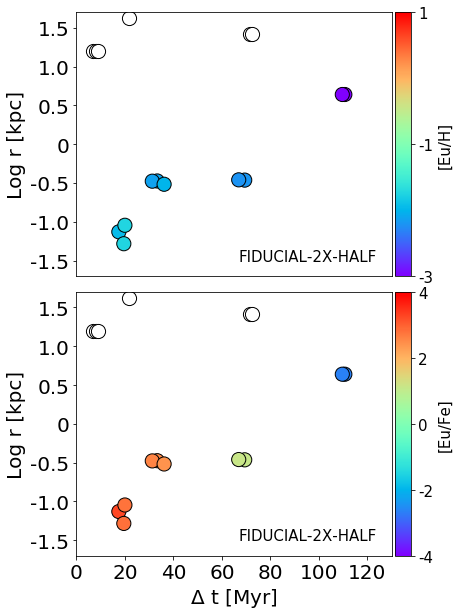} }}%
    \caption{The distance between the NSM site and newly formed stars for each selected run vs. time since the merger (same runs as in Fig.~6). The color coding indicates the europium abundances of stars formed after the NSM event. Stars that are too far away to contain europium abundance are denoted by open circles. We find that highly Eu-enhanced stars ($\rm [Eu/Fe]>2$) only form near the NSM site within $\sim$300 pc. We show that the [Eu/H] and [Eu/Fe] ratios decrease with increasing distance from the NSM site, but Eu-enhanced stars continue to form until $\sim$\,60~Myr after the NSM event. Given that the run with the Eu yield reduced by a half (right panels) still produces stars with high Eu abundances, we conclude that the most important factor determining the formation of highly Eu-enhanced stars is how quickly new stars can form in the gas cloud where the NSM occurs.}
\label{fig:example}%
\end{figure*}

\subsubsection{Single galaxy comparison runs}

As a second approach, we perform several comparison simulations of a single UFD, varying key NSM model parameters. To increase the frequency of NSMs, we allow one NSM event to take place per $M_{\ast} \approx 3.5\times10^3\msun$. Although this frequency is higher than the empirical estimate by an order of magnitude, it assures the simulated UFD to experience at least one NSM event during its SFH. We carry out additional simulations by extending the period (or decreasing the frequency) of NSM events by a factor of $\sim$2 and $\sim$3, compared to the fiducial value, such that a single NSM occurs per $7\times10^3\msun$ or $1\times10^4\msun$ in stars, respectively. We also study the effect of reducing the europium yield by half, $M_{\rm Eu} \approx 2\times10^{-5}\msun$, to achieve a value similar to the one inferred from GW170817.

We note that, for the purpose of the r-process study, in "Single galaxy" runs we terminate the simulations immediately after reionization due to computational cost. Any derived r-process abundances presented here may thus not reflect the final sample for a given UFD analog at $z=0$. In Figure~6, we compare the resulting europium abundances for the simulated UFDs with the values observed in Ret~II. Although we set up the four simulations so that all should experience a NSM event, only two runs ({\sc fiducial-2x} and {\sc fiducial-2x-half}) contain stars with highly enhanced europium abundances. In the runs where we double the NSM period, or additionally reduce the Eu yield by half, the resulting [Eu/H] and [Eu/Fe] values at [Fe/H]$\simeq -3.5{\rm \ to} -2$ are lower than what is observed for the stars in Ret~II. The high values in {\sc fiducal-2x-half} for [Eu/Fe] at [Fe/H]$\lesssim-3.5$, where no counterparts are observed, are mainly due to the small amount of Fe, resulting in high [Eu/Fe].

To investigate how far the r-process elements, ejected from the NSM, could reach and pollute the surrounding gas, we display in Figure~7 the europium abundances (encoded in color) of the simulated stars as a function of the time elapsed after the NSM event (for the same runs as in Fig.~6). We distinguish these histories for different distances between the NSM site and the location of the newly formed stars. The open circles indicate the stars showing a lack of europium abundance because they are far away from the NSM site. It is evident that stars with relatively high europium values, [Eu/H]$\,\gtrsim-2$ or [Eu/Fe]$\,\gtrsim2$, formed close to the NSM site, within $\sim$300~pc. Both panels demonstrate that the Eu abundances decrease with increasing distance from the NSM site. However, the gas close to the NSM site can maintain such high europium values for a long period of time, $80-100$~Myr, giving rise to r-process enhanced stars. Considering that the Eu-enhanced stars still emerge in the run with the yield reduced by half (right panels), it appears that the most important determining factor in forming such highly Eu-enhanced stars is how quickly new stars can form around the merger site, rather than the Eu yield. Put differently, the key here is how quickly the gas, altered by stellar feedback, could recombine and recollapse, reaching the high density required to form subsequent stars.

To further study this, in Figure~8 we show the hydrogen number density (top panels) and europium abundance (bottom panels) of the gas as a function of distance from the NSM site about 10 (left), 50 (middle), and 100 (right) Myr after the NSM event for the {\sc fiducial-2x-half} run (same as the right panels of Fig.~6). As can be seen, about 10~Myr after the NSM (left panels), the surrounding gas begins to be contaminated by the ejected europium. Even though the gas density in the inner region, $r\lesssim$50~pc, is lowered by stellar feedback, the density peaks at $r<$300~pc are high enough to form stars with [Eu/H]$\approx-2$. Until 100~Myr after the onset of the NSM event, the evacuated gas continues to fall back to the central region, so that next generation stars can form out of this highly enriched dense gas cloud. The [Eu/H] ratios of those high density peaks decline from [Eu/H]$\approx0$ at $\Delta t \approx 10$~Myr to [Eu/H]$\approx -1$ ($r\approx$10 pc) and [Eu/H]$\approx -2$ ($r\approx$ 40 pc) at $\Delta t \approx100$~Myr. As illustrated in Figure~8, multiple gas density peaks could exist in close proximity to the NSM site, such that a number of r-process enhanced stars could form within those high-density clouds.

One of the main obstacles for producing r-process enhanced stars in the UFD analogs is the short duration of their star formation histories. We find that all the simulated UFDs stop forming stars at the epoch of reionization. Consequently, if a NSM takes place just before the onset of reionization, with insufficient time for the enriched gas to reach high density, no r-process enhanced stars will form. When we double the interval between NSM events, such that one NSM event occurs per $7\times10^3\msun$, stars with high europium abundance can still form. However, in the run with three times longer intervals ({\sc fiducial-3x}), no high europium stars are found, because the NSM happens right before reionization. As explained previously, the amount of europium produced does not seem to be the main factor, because Eu-enhanced stars can still form, even if the europium yield decreases by a half. Therefore, the decisive factor in forming r-process enhanced stars is whether or not the stars can form in the close spatial and temporal vicinity of the NSM event.

\begin{figure*}
    \centering
    \subfigure{{\includegraphics[width=5cm]{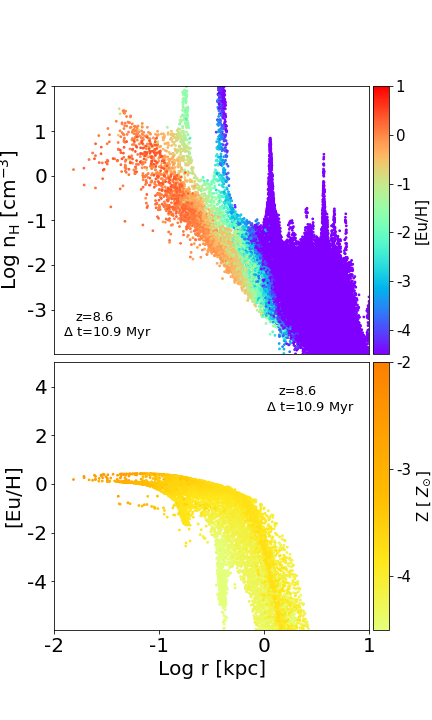} }}%
    \subfigure{{\includegraphics[width=5cm]{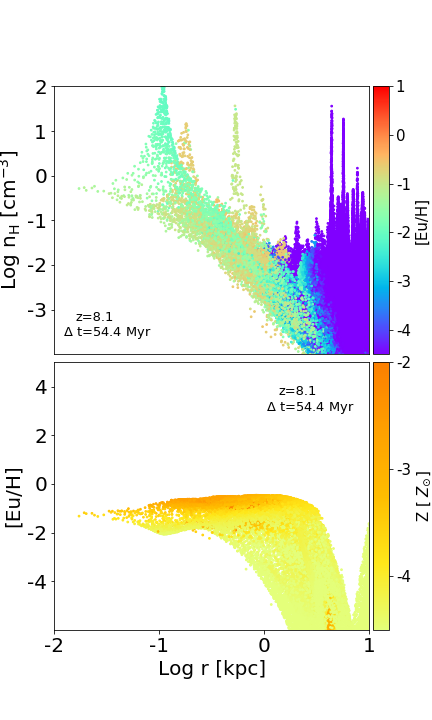} }}%
        \subfigure{{\includegraphics[width=5cm]{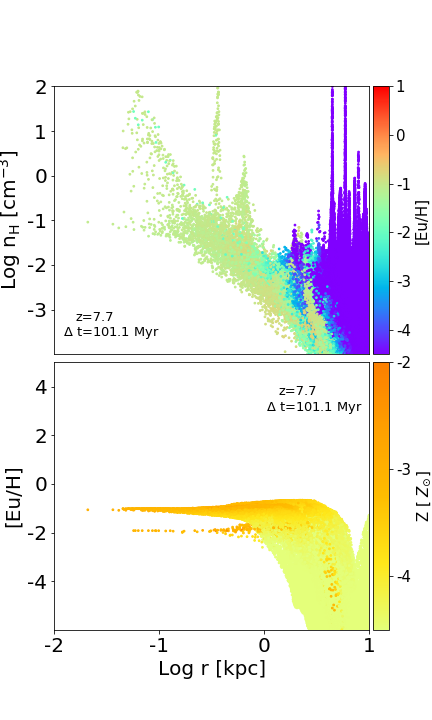} }}%
    \caption{The hydrogen number density (top panels) and europium abundance (bottom panels) of the gas as a function of distance from the NSM site at 10 (left), 50 (middle), and 100 (right) Myr after the NSM event. We show the results for the run, where we increase the interval between NSM events and reduce the europium yield by a half ({\sc fiducial-2x-half}), corresponding to the right column in Fig.~6. About $\sim$ 10~Myr after the NSM, hydrogen number densities within 50~pc have decreased from $n_{\rm H}\gtrsim100$ $\rm cm^{-3}$ to $n_{\rm H}\lesssim1$ $\rm cm^{-3}$, whereas the densest regions with high Eu abundance ([Eu/H]>-2.0) are located at $<300$~pc, where new Pop~II stars form. At $\sim 50$~Myr after the NSM event, the central region is replenished by Eu-enriched gas to a level of [Eu/H]$\approx-2$, continuously forming stars.}%
    \label{fig:example}%
\end{figure*}

\subsection{Comparison with other work}

Numerous theoretical arguments have been made to account for the observed r-process enhanced stars within the assembly process of the MW halo (e.g. \citealp{VandeVoort2015, Shen2015, Naiman2018, Safarzadeh2019, vandeVoort2020, Dvorkin2020, Tarumi2021}), dSphs (e.g. \citealp{Hirai2015}), or UFD galaxies (e.g. \citealp{Safarzadeh2017, Beniamini2016b, Beniamini2018, Brauer2019, Tarumi2020}), thus constraining the r-process site. Note that in the case of the MW halo, the main focus of including r-process sites is to explain the global features, i.e. the relation of $\rm [Eu/Fe]-[Fe/H]$, observed in the disk stars, rather than examining the existence of highly r-process enhanced stars.

Regarding r-enhanced stars in UFDs, \citet{Safarzadeh2017} have reached similar conclusions to our work, in that the emergence of the highly r-enhanced stars in Ret~II is strongly subject to galactic environment, i.e. where the NSM event happens: if the event occurs in the center of a UFD-like galaxy, as opposed to off-center, Ret~II-like stars can be produced, as r-process ejecta are less dispersed. The importance of the local environment of the NSM site, combined with a prolonged SFH over a few 100 Myr, is also emphasized in the recent work by \citet{Tarumi2020}, suggesting that the r-process abundance patterns in Ret~II and Tuc~III could be reproduced by a NSM emerging inside or around the virial radius of a UFD progenitor.

The off-center NSMs, in particular, are realized by natal kicks to neutron stars due to the explosion of their progenitor stars (e.g. \citealp{Tauris2017}), which might place a NSM event at the outskirt of a galaxy, thus lowering the possibility of forming highly Eu-enhanced stars. We note that our derived stellar [Eu/H] values should be considered as upper limits, given that we did not consider natal NSM kicks. A weak natal kick might be a prerequisite for neutron stars to be bound in UFDs, and especially to explain Ret~II-like systems. \citet{Beniamini2016b} have demonstrated that a significant population of neutron stars with weak natal kicks (\citealp{Beniamini2016a}) could have a center of mass velocity less than 15 $\rm km/s$, comparable to the escape velocity of small galaxies such as UFDs.

One of the difficulties in creating r-process enhanced stars in UFDs is that the delay time between the formation of neutron stars and their final merger needs to be shorter than the duration of star formation activity in UFDs. However, for UFDs, their star formation activity lasts of order a few hundred million years, being ultimately truncated by reionization and stellar feedback (e.g. \citealp{Jeon2017}). Note that the duration of star formation of the simulated galaxies in this work is also a few 100 Myr. A precise delay time can be calculated from population synthesis models for binary systems, resulting in a delay time distribution (DTD), with a typical form of $t_{\rm merger}\propto t^{-1}$ (e.g. \citealp{Dominik2012, Mennekens2016}), ranging from a minimum, $t_{\rm min}$, to a maximum time scale, $t_{\rm max}$. Our adopted delay time of $\sim5$ Myr in this work corresponds to a fast merging channel (e.g. \citealp{Belczynski2001, DeDonder2004, Dominik2012}). Such rapid merging might correspond to the extreme end of the DTD, given that the delay time, inferred from GW170817, exceeds 3-4~Gyr (e.g. \citealp{Pan2017, Skuladottir2020}). Other theoretical studies seem to support rapid merging to reproduce Ret~II-like systems. For instance, using the fiducial power-law DTD ($t^{-1}$), \citet{Safarzadeh2019} have shown that even for a minimum merging time of $t_{\rm min}=1$~Myr, corresponding to a median timescale of a few 10~Myr, the resulting r-process abundances turn out to be an order of magnitude lower than what is observed both in the MW halo and Ret~II.

Although a NSM event is a compelling r-process site for UFDs, given that r-process enhanced stars in UFDs might originate from a very rare and prolific event, rare SNe could be an alternative r-process site. Especially when considering r-process enhanced stars in the MW, there is an ongoing debate on their origin (e.g. \citealp{vandeVoort2020, Banerjee2020, Dvorkin2020}). For example, \citet{Cote2019} have shown that NSMs alone may be insufficient to match observed Eu-abundances, in particular, the decreasing trend of $\rm [Eu/Fe]-[Fe/H]$ at $\rm [Fe/H]>-1$ in MW disk stars, suggesting a rare class of SNe as the origin of r-process elements.

Lastly, it is worth discussing the relationship between reionization and the possibility of forming Ret~II-like systems. The onset of reionization is an important factor in determining how long star formation can be sustained in dwarf galaxies (e.g. \citealp{Simpson2013, Jeon2017}). Although in this work the onset of reionization is designed to occur at the same time in all the simulated UFDs, in reality, satellite galaxies undergo different reionization processes depending on their distance from a host galaxy.

It has been suggested that Ret~II could be a satellite of the Large Magellanic Cloud (LMC), in which case Ret~II might have existed in a more isolated environment at the time of reionization (e.g. \citealp{Sales2017, Kallivayalil2018, Nico2020, Patel2020}). Particularly, \citet{Patel2020} reinforced the possibility of Ret~II being a satellite of the LMC based on the derived orbital history using the recently measured proper motion data from the Gaia DR2 (e.g. \citealp{Gaia2018}). If Ret~II existed in an isolated environment, reionization might have been delayed, making the progenitor of Ret~II sustain star formation longer and thus increasing the possibility of forming MP r-II stars.

Not only when reionization started, but also when it ended is important. Recent observations propose that reionization might last up to $z\approx5.5-6$ (e.g. \citealp{Becker2015, Choudhury2015, McGreer2015, Puchwein2019}). Such a late reionization scenario could increase the possibility of NSM events occurring in the progenitors of UFD systems, and thus the possibility of the Eu-polluted star formation as the quenching of stars could also be delayed.

\section{Summary and Conclusions}

We have investigated the conditions under which extremely r-process enhanced metal poor stars, the so-called MP~r-II ($\rm [Eu/Fe] > 1, [Fe/H] < -1.5$) stars, as observed in Ret~II, can arise in UFDs. To do this, we have performed cosmological hydrodynamic zoom-in simulations of low-mass halos ($M_{\rm vir}\approx10^7-10^8\msun, M_{\ast}\lesssim10^{5}$ at $z=0$), which can be considered as direct analogs of UFDs in the local Universe. Given that most of the stars in observed UFDs are composed of ancient populations, it is expected that star formation is likely to be quenched in the early Universe, both by reionization and SN feedback (e.g. \citealp{Brown2014, Weisz2014, Jeon2017}), resulting in relatively short SFHs, a few hundred Myr.

In this regard, the existence of MP r-II stars in the MW UFDs, such as Ret~II, has been puzzling because the NSM event, one of the most promising pathways of producing r-process elements, is necessarily rare in systems with such short SFHs. In this work, we scrutinize whether MP r-II stars can be self-consistently produced in cosmological zoom-in simulations that employ a simple toy model for NSM events. This model is described by two parameters: the NSM event rate and europium yield. First, we have conducted relatively large volume zoom-in simulations, {\sc LV1} and {\sc LV2}, in which multiple UFDs emerge, to ascertain if any Ret~II-like system could arise, when adopting parameters consistent with observations. We then have focused on a single galaxy to explore the environmental conditions for the generation of MP r-II stars, by changing the NSM parameters. The main findings from this study are as follows.

\begin{itemize}

\item 
We confirm that the global properties of the simulated UFD analogs, such as half-stellar mass radius, average stellar metallicity, and stellar mass, show good agreement with observational data.
\\
\item {Large volume runs: \sc LV1 and LV2}
\begin{itemize} 
\item
Adopting a simple toy model for NSMs, whose parameters are in line with those inferred from observations such as the NSM rate of a single NSM per $10^5 \msun$ and the europium yield of $M_{\rm Eu}\approx4.5\times10^{-5}\msun$, we identify only one Ret~II-like system ($\rm M_{vir}\approx4.6\times10^7\msun$, $M_{\ast}\approx 3.4\times 10^3\msun$ at $z=0$) out of $\sim32$ UFD analogs in {\sc LV2}. Meanwhile, no Ret~II-like system is found in {\sc LV1}, where $\sim8$ UFD analogs emerge. This frequency is largely consistent with the observation that only 1 in 10 UFDs (Ret~II) contains highly r-process enhanced stars (e.g. \citealp{Ji2016a}).
\\
\item
The derived europium abundances of MP r-II stars in the simulated Ret~II-like galaxy are $\rm [Eu/Fe] > 2.5$ and $\rm [Eu/H]>-0.6$ at $\rm [Fe/H] < -2.0$, similar to the abundances observed in Ret~II. 
\\
\item
Unlike Ret~II, in which three MP r-II stars are CEMP stars, no CEMP stars are found in the simulated Ret~II-like system despite the inclusion of Pop~III stars. Considering the occurrence of CEMP stars at the average rate of $40\%$ in the other 31 haloes in {\sc LV2}, the lack of CEMP stars in the Ret-II like halo is simply due to a stochastic effect. The $\alpha$-element abundances, on the other hand, are consistent with what is observed in Ret~II.
\end{itemize}
\item {\sc Single galaxy comparison runs}
\begin{itemize}
\item
Building upon a fiducial single galaxy run, which adopts a rate of one NSM per $3.5\times10^3\msun$ and the europium yield of $M_{\rm Eu}\approx4.5\times10^{-5}\msun$, we have conducted an additional 3 simulations to explore whether MP r~II stars can be reproduced by decreasing the NSM rate by a factor of 2 and 3, and additionally reducing the Eu yield by half. We discover MP r~II stars ($\rm [Eu/Fe]>1$, $\rm [Fe/H]\lesssim-1.5$) in the two runs, the case where the NSM rate is increased by a factor of 2 ({\sc Fiducial-2x}) and where the Eu yield is reduced by half ({\sc Fiducial-2x-half}).
\\
\item
We suggest that the most important factor in determining whether such highly r-enhanced stars can form is how quickly subsequent star formation can resume in gas clouds that experienced such NSM events. Although r-process elements are diluted with increasing distance from a NSM site, we find that the gas near the merger site can maintain europium values high enough to produce MP r-II stars for a prolonged period of 80-100~Myr.
\\
\item
We find that NSM events must occur long before (a few 10-100 Myr) the onset of reionization to create MP r-II stars in UFDs. Otherwise, the gas evacuated out of the host halo by stellar feedback might not have enough time to be replenished, to trigger subsequent star formation.
\end{itemize}
\end{itemize}

As discussed in the previous section, our NSM toy model is highly idealized, assuming fast merging and not taking into account natal kicks. Nevertheless, given the short SFH of low-mass UFDs ($<$1 Gyr), it is evident that the time interval between the NSM event and subsequent star formation ought to be shorter than the SFH timescale of UFDs. This constraint, however, could be alleviated if relatively massive dwarf galaxies, experiencing NSM events during their long SFHs, subsequently lose mass as they undergo tidal stripping. Such scenario (e.g. \citealp{Penarrubia2008, Hansen2017}) is one of the most promising pathways to account for the MP r-I stars in Tuc~III, which seems to have lost most of its stellar mass, resulting in the extremely low $M_{\ast}\approx800\msun$ (e.g. \citealp{Drlica2015}).

Understanding the formation of r-process enhanced stars in the most ancient and least massive galaxies in the Universe addresses not only the NSM event itself, but also the complex interplay between early stellar populations, galaxy assembly, metal mixing, and nuclear physics. Additional discoveries in UFDs, exploring r-process elements, will be necessary to better understand the physics of the r-process sites. Future observations with greatly improved spectroscopic sensitivity, such as provided by the Giant Magellan Telescope (GMT), will elucidate the detailed chemical abundance patterns left behind by the r-process site, thus helping to distinguish between different r-process models (e.g. \citealp{Ji2018}). Similar to the CEMP signature, the early Universe may provide an ideal laboratory to probe the elusive r-process as well.

\section*{acknowledgements}
We are grateful to Volker Springel, Joop Schaye, and Claudio Dalla
Vecchia for permission to use their versions of \textsc{gadget}. We thank the anonymous referee for the constructive comments which helped to improve the manuscript. G.~B. acknowledges support from HST Grant 15030. The simulations were performed using the Nurion Super computer at the KISTI National Supercomputing Center through Grant No. KSC-2020-CRE-0175. We utilized \textsc{yt} for data visualization and analysis tools. M.~J. is supported by a grant from Kyung Hee University in 2020 (KHU-20201115).

\section*{DATA AVAILABILITY}
The simulation data and results of this paper may be available upon
request.

\footnotesize{

\bibliography{myrefs2}{}
\bibliographystyle{mnras} 








\bsp	
\label{lastpage}
\end{document}
